\documentclass{article}
\usepackage{amsmath}
\usepackage{amscd}
\usepackage{amsthm}
\usepackage{amssymb} \usepackage{latexsym}
\usepackage{eufrak}
\usepackage{euscript}
\usepackage{epsfig}
\usepackage{graphics}
\usepackage{array}
\usepackage{enumerate}

\theoremstyle{theorem}

\theoremstyle{corollary}

\theoremstyle{lemma}

\theoremstyle{definition}

\theoremstyle{proof}

\theoremstyle{remark}

\newcommand{\bel}[1]{\begin{equation}\label{#1}}

\newcommand{\be}{\begin{equation}}

\newcommand{\ba}{\begin{eqnarray}}
\newcommand{\ea}{\end{eqnarray}}
\newcommand{\rf}[1]{(\ref{#1})}
\newcommand{\bi}{\bibitem}
\newcommand{\qe}{\end{equation}}

\begin{document}
\title{Graph spectra as a systematic tool in computational biology}
\author{Anirban Banerjee, J\"urgen Jost\thanks{This author thanks the local organizers B\"ulent
  Karas\"ozen and Gerhard Wilhelm Weber for an intellectually
  stimulating meeting and outstanding hospitality.}\\[2ex]
 {\small Max Planck Institute for Mathematics in the
  Sciences,}\\ 
{\small Inselstr.22, 04103 Leipzig, Germany,}\\
 {\tiny banerjee@mis.mpg.de,
  jost@mis.mpg.de}}
\maketitle
\begin{abstract}
 We present the spectrum of the (normalized) graph Laplacian as a
 systematic tool for the investigation of networks, and we describe
 basic properties of eigenvalues and eigenfunctions. 
Processes of graph formation like motif joining or duplication  leave characteristic traces in the
spectrum. This can suggest  hypotheses about the evolution of a graph representing
 biological data. To this data, we analyze several biological networks
 in terms of rough qualitative data of their spectra.
\end{abstract}

\bigskip

\section{Introduction}
Modern biological data are often represented in terms of
graphs. Microarray data may lead to graphs whose vertices are genes
and whose edges stand for correlations, hypothetically interpreted as
interactions. In the study of the proteome of a cell, one sees
protein-protein interaction networks. Likewise, at a higher level,
cell-cell interactions naturally lead to interaction graphs. A
particular example are neural networks where the vertices stand for neurons
and the edges for synaptic connections. In populations, graphs encode
networks of interactions between individuals, and in ecosystems,
trophic and other 
interactions between species. A special case are phylogenetic trees
that express descendence relations between species. \\
The natural question then is how biological content can be extracted
from these formal structures, the graphs to which the biological data
are reduced. In graph theory, many concepts have been developed that
capture various quantitative or qualitative aspects of a
graph (for an algebraic, graph theoretical approach, see
e.g. \cite{GoRo,Bol}, for statistical mechanics methods see e.g. \cite{AB,New,DoMe}). Recently, a power law behavior of the degrees has become quite
popular as it seems to be rather ubiquitous in biological and other
data (\cite{BA}). Another powerful invariant of the graph is its first eigenvalue
that provides estimates for how difficult it is to cut up the graph
into disjoint components (see \cite{Chung}, or for how easily dynamics at the vertices
can get synchronized (\cite{PeCa,PRK,JJ1,AJW}) and many other articles).\\
Useful as any such individual invariant may be, however, it cannot
capture all the qualitative aspects of a graph. For example, graphs
with the same degree distribution can have a completely different
synchronizability (\cite{ABJ1,ABJ2}). Also, by their very nature, universal properties
like a power law degree distribution  capture what
is common to large classes of graphs, but fail to identify what is
specific about graphs from a particular domain, and what distinguishes
those graphs qualitatively from those from other fields. \\
Therefore, in this contribution, we advocate a set of graph invariants
that, on
one hand,  can be easily graphically represented and therefore visually
analysed and compared, and on the other hand, yields an essentially
complete qualitative characterization of a graph. This is the spectrum
of the graph Laplacian (\cite{Merris,Mohar,J1,J2,ZW,BiLeSt,BJ2}). In \cite{BJ1}, we have applied this method to
the study of protein-protein interaction networks. \\

\section{The spectrum of a graph}
Let $\Gamma$ be a finite and connected graph with $N$ vertices. Vertices $i,j
\in\Gamma$ that are connected by
an edge of $\Gamma$ are called neighbors, $i\sim j$. The number of 
neighbors of a vertex $i \in
\Gamma$ is called  its degree $n_i$. For functions $v$ from the vertices of $\Gamma$ to
$\mathbb{R}$, we define the (normalized) Laplacian as
\bel{3}
\Delta v(i):=  v(i) -\frac{1}{n_i} \sum_{j, j \sim i}v(j) .
\end{equation}
(Note that this operator is different  from, and in particular, has a
  different spectrum than the operator  $Lv(i):=n_i v(i)-\sum_{j, j \sim i}v(j)$
  usually   studied in the graph theoretical literature as the
  (algebraic) graph Laplacian, see e.g. \cite{Bol,GoRo,Merris,Mohar,BiLeSt}, but
  has the same spectrum as the Laplacian investigated in \cite{Chung}. The
  normalized Laplacian is the operator underlying random
  walks on graphs, and it 
  naturally incorporates a conservation law.)\\

We are interested in the spectrum of this operator as yielding
important invariants of the underlying graph $\Gamma$ and
incorporating its qualitative properties. As in the case of the algebraic
Laplacian, one can essentially recover the graph from its spectrum, up
to isospectral graphs. The latter are known to exist, but are --
arguably\footnote{For example, most trees are not  uniquely
  determined by their spectrum.} -- 
relatively rare and qualitatively quite similar in most respects (see
\cite{ZW} for a survey). For
a heuristic algorithm for recovering a graph from the spectrum of its algebraic
    Laplacian which can be easily
modified for the normalized Laplacian, see \cite{IpMi}.\\
We now recall some elementary properties, see e.g. \cite{Chung,JJ1}. The normalized Laplacian, henceforth simply called the Laplacian, is
symmetric for the product
\bel{2}
(u,v):=\sum_{i \in V} n_i u(i)v(i)
\end{equation}
for real valued functions $u,v$ on the vertices of $\Gamma$. $\Delta$ is nonnegative in the sense that
$(\Delta u,u) \ge 0$ for all $u$. \\
The eigenvalues of $\Delta$ therefore 
are real and nonnegative, 
  the eigenvalue equation being
\bel{6}
\Delta u - \lambda u =0.
\end{equation}
A nonzero solution $u$ is called an eigenfunction for the eigenvalue
$\lambda$. Since $\Gamma$ has $N$ vertices, the function space on
which $\Delta$ operates is $N$-dimensional. Therefore, it has $N$
eigenvalues; some of them  might occur with multiplicity $>1$. The
eigenfunctions for an eigenvalue $\lambda$ constitute a vector space
whose dimension equals the multiplicity of the eigenvalue
$\lambda$. In the sequel, when we describe an eigenfunction, this is
to be taken as some suitable element of this vector space.\\
The smallest eigenvalue is $\lambda_0 =0$, with a constant
eigenfunction.  This eigenvalue is simple because we
  assume that $\Gamma$ is connected; in general, the multiplicity of
  the eigenvalue 0 equals the number of connected components, with the
  corresponding eigenfunctions being $\equiv 1$ on one and $\equiv 0$
  on all other components. Returning to our case of a connected graph
  $\Gamma$, then
\bel{7}
\lambda_k >0
\end{equation}
for $k>0$ where we order the eigenvalues as
$$ \lambda_0=0 < \lambda_1 \le ... \le \lambda_{N-1}.$$
After the brief discussion of the smallest eigenvalue, 0, we now turn
to the largest one; here, we have
\bel{8}
\lambda_{N-1} \le 2,
\qe
with equality iff the graph is bipartite. Thus, a single eigenvalue
determines the global property of bipartiteness. In fact, it is also
true that a graph is bipartite iff
 whenever $\lambda$ is an eigenvalue, then so is
$2-\lambda$. In other words, a bipartite graph has a spectrum that is
symmetric about 1, and this characterizes bipartiteness.\\
It is also instructive to look at the spectrum of particular
graphs. For example, for a complete graph of $N$ vertices, we have
\bel{9}
\lambda_1 = ... = \lambda_{N-1}=\frac{N}{N-1},
\qe
that is, the eigenvalue $\frac{N}{N-1}$ occurs with multiplicity
$N-1$. Among all graphs with $N$ vertices, this is the largest
possible value for $\lambda_1$ and the smallest possible value for
$\lambda_{N-1}$. Again, this spectral property fully characterizes complete
graphs.\\
The preceding examples concern exact values for the eigenvalues. In
contrast, qualitative properties of a graph are usually characterized
by inequalities for its eigenvalues, an issue that we shall return to
below.

\section{Eigenfunctions}
When we think of a graph $\Gamma$ representing biological data as a structure
that has evolved from some simpler precursors, for example by joining
smaller graphs into a larger one, or by duplicating certain sets of
vertices in a precursor graph, it is important to find some
indications of this process in the spectrum of $\Gamma$. It turns out
that also certain properties of eigenfunctions can be useful here. We
shall now describe some such aspects in formal terms (for some
details, we refer to \cite{BJ2}).\\
In some cases, a solution $u_k$ of   the eigenvalue equation 
\bel{10}
\Delta u_k - \lambda_k u_k =0
\end{equation}
can be localized, that is, be 0 outside a small set of vertices. In
other cases, it has to be 
global, that is, be 0 only at relatively few vertices. These are
qualitative notions, but they provide some insight into the behavior
of graphs under certain operations as we shall now explore. \\
The considerations will depend on
the eigenvalue equation \rf{6}, rewritten as
\bel{11}
\frac{1}{n_i}\sum_{j \sim i} u(j)=(1-\lambda)u(i) \text{ for all }i.
\qe
We observe that when the eigenfunction $u$ vanishes at $i$, then also
\bel{12}
\sum_{j \sim i} u(j)=0.
\qe 
The converse also holds, except for the case $\lambda=1$ when \rf{12}
holds at all points regardless of whether the eigenfunction vanishes
there or not.\\
We start with constructions that lead to localized eigenfunctions. We
think of a motif as a small graph whereas the graph $\Gamma$ is
supposed to be large. This is not at all necessary for the subsequent
constructions, but is in the spirit of the term ``localized''.
\begin{enumerate}
\item {\bf Motif joining:}  Let $\Gamma_0$ be another graph, $j_0$ a
  vertex of $\Gamma_0$, with an
  eigenvalue $\lambda$ and an eigenfunction $u^\lambda$ that vanishes
  at $j_0$, i.e., $u^\lambda(j_0)=0$. When we then form a graph
  $\bar{\Gamma}$ by identifying the vertex $j_0$ with an arbitrary
  vertex $i$ of $\Gamma$, the new graph $\Gamma_0$ also has the
  eigenvalue $\lambda$, with an eigenfunction that agrees with
  $u^\lambda$ on $\Gamma_0$ and vanishes at the other vertices, that
  is, those coming from $\Gamma$. Thus, a motif $\Gamma_0$ can be
  joined to an existing graph with a preserved eigenvalue and a
  localized eigenfunction when the joining occurs at one (or several)
  vertices where that eigenfunction vanishes. 
\item {\bf Motif duplication:} Let  $\Gamma_1$ be a motif in $\Gamma$,
  that is, a (small) subgraph of $\Gamma$, with vertices $j_1,\dots ,j_m$. Let the function $u$ on the
  vertex set of $\Gamma_0$ satisfy
\bel{21}
\frac{1}{n_i}\sum_{j\in \Gamma_1, j \sim i} u(j)=(1-\lambda)u(i) \text{
  for all }i\in \Gamma_1 \text{ and some } \lambda.
\qe
Let $\bar{\Gamma}$ be obtained from $\Gamma$ by doubling the motif
$\Gamma_1$, that is, by adding vertices
$i_1,\dots ,i_m$ and their connections as in $\Gamma_1$ and connecting each $i_\alpha$ with all $i \notin
\Gamma_1$ that are neighbors of $j_\alpha$. Then the  graph
$\bar{\Gamma}$ possesses the eigenvalue $\lambda$ with an eigenfunction
$u^\lambda$ that is localized on $\Gamma_1$ and its double; it agrees with $u$
on $\Gamma_1$, with $-u$ on the double of $\Gamma_1$, and vanishes on the
rest of $\bar{\Gamma}$. Thus, the eigenvalue $\lambda$ is produced
from motif duplication with symmetric eigenfunction balancing. 
\end{enumerate}

Not all eigenvalues possess localized eigenfunctions. 
Take  cyclic graphs $\Gamma_1, \Gamma_2$ of lengths
$4m-1$ and $4n +1$, for some positive integers $m,n$. Since the only
cyclic graphs that admit the eigenvalue 1 are those of length $4k$,
neither $\Gamma_1$ nor $\Gamma_2$ possesses the eigenvalue 1, but
if we join them by identifying a vertex $i_0\in \Gamma_1$
with a vertex $j_0\in \Gamma_2$  the resulting graph $\Gamma$ has 1 as an
eigenvalue. An eigenfunction has the value 1 at the joined vertex, and
the values $\pm 1$ occurring always in neighboring pairs at the other vertices of $\Gamma_1, \Gamma_2$, where the two neighbors of $i_0=j_0$ in $\Gamma_1$
both get the value $-1$, and the ones in $\Gamma_2$ the value
1. Since the multiplicity of the eigenvalue 1 on $\Gamma$ is 1, there
exists no other linearly independent eigenfunction for the eigenvalue
1. Thus, the local construction of joining two graphs at a single
vertex here produces an eigenfunction that cannot be localized. \\
As another example, we can take any two graphs
$\Gamma_1,\Gamma_2$. Their disjoint union then has two components, and
therefore, the multiplicity of the eigenvalue 0 is 2. One
eigenfunction $u_0$ is $\equiv 1$ on $\Gamma_1$ and $\equiv 0$ on
$\Gamma_2$, and for the other one, $v_0$, the roles of the components are
reversed. When we now form a graph $\Gamma$ by connecting some vertex $i_0 \in\Gamma_1$ to some
vertex $j_0 \in \Gamma_2$ by an edge, then the multiplicity of the
eigenvalue $\lambda_0=0$ becomes 1 because $\Gamma$ is
connected; the corresponding eigenfunction $u$ is $\equiv 1$. However, when both $\Gamma_1$ and $\Gamma_2$ are large, the
next\footnote{assuming for simplicity that $\Gamma_1,\Gamma_2$ do not
  have small nontrivial eigenvalues themselves} eigenvalue $\lambda_1$ of $\Gamma$ is very small, and a
corresponding eigenfunction is well approximated by one, $v$, that equals a
positive constant on $\Gamma_1$ and a negative constant on $\Gamma_2$
(satisfying $\sum_{i\in \Gamma} n_i v(i)=0$). Thus, $u$ is a symmetric
linear combination, $v$ an antisymmetric one of the original
eigenfunctions $u_0,v_0$, and also the eigenvalues are close.

\section{Properties of spectral plots and evolution hypotheses}
Constructions like motif joining or duplication describe certain
processes of graph formation that leave characteristic traces in the
spectrum. This suggests that they can also serve useful roles for
developing hypotheses about the evolution of a graph representing
actual biological data. Of course, such hypotheses then need to be
biologically plausible as well. Let us consider some examples. The
simplest version of motif duplication is the doubling of a single vertex
$j_1\in \Gamma$. According to the general scheme, we add a new vertex $i_0$
and connect $i_0$ with all neighbors of $j_0$. This generates an
eigenvalue 1, with an eigenfunction $u_1$ that is localized at $j_0$
and $i_0$, $u_1(j_0)=1,u_1(i_0)=-1$. Thus, if the spectral plot of a
graph has a high peak at the eigenvalue 1, a natural hypothesis is
that this graph evolved via a sequence of vertex doubling.\\
The next simplest case of a motif is an edge connecting two vertices $j_1,j_2$. \rf{21} then becomes
\bel{31}
\frac{1}{n_{j_1}}u(j_2)=(1-\lambda)u(j_1),\quad
\frac{1}{n_{j_2}}u(j_1)=(1-\lambda)u(j_2),
\qe
with the solutions 
\bel{24}
\lambda_{\pm}=1\pm \frac{1}{\sqrt{n_{j_1}n_{j_2}}}.
\qe
Thus, the duplication of an edge produces the eigenvalues
$\lambda_{\pm}$. These are symmetric about 1.  Also, when the
degree of $j_1$ or $j_2$ is large, $\lambda_{\pm}$ are close to
1. Thus, when the spectral peak at 1 is  high, but not too sharp, and
symmetric about 1, this
is  an indication that edge duplication has played some role in
the evolution of the structure.\\
Next,  we connect an edge between vertices $j_1,j_2$ to an
existing graph $\Gamma$ by connecting both $j_1$ and $j_2$ via an edge
to some vertex $i_0\in \Gamma$, or equivalently, we join a triangle
with vertices $j_0,j_1,j_2$ to $\Gamma$ by identifying $j_0$ with
$i_0\in \Gamma$. In that case, we produce the  eigenvalue
3/2. An eigenfunction $u$ for the eigenvalue 3/2
satisfies $u(j_1)=1,u(j_2)=-1$,  and  vanishes
elsewhere. Thus, again, it is localized. The same result obtains when
we join the triangle by connecting $j_0$ and $i_0$ by an edge instead
of identifying them. A high
multiplicity of the eigenvalue 3/2 may then generate the hypothesis that
such  processes of triangle joining repeatedly occurred in the evolution of the
structure.\\
When in addition to the triangle $j_0,j_1,j_2$ another triangle
$k_0,k_1,k_2$ is joined by identifying both $j_0$ and $k_0$ with
$i_0\in \Gamma$, we not only generate the eigenvalue 3/2 with
multiplicity 2, but also the eigenvalue 1/2, with an eigenfunction 
$v(j_1)=v(j_2)=1, v(k_1)=v(k_2)=-1$ and 0 elsewhere. Again, such a
feature when prominently observed in a spectral plot may induce a corresponding
hypothesis. \\
The described operations can also be of a global nature. For example,
we can double the entire graph $\Gamma$; when $\Gamma$ consists of the
vertices $p_1,\dots ,p_N$, we take another copy $\Gamma'$ with
vertices $q_1,\dots ,q_N$ and the same connection pattern and connect
each $q_\alpha$ also to all neighbors of $p_\alpha$. The new graph
$\bar{\Gamma}$ then has the same eigenvalues as $\Gamma$, plus the
eigenvalue 1 with multiplicity $N$. This is biologically relevant,
because there is  some evidence for whole genome duplication
\cite{Ohno1970, Wagner1994, WolfeShields1997}. However,
protein-protein interaction networks do have a high multiplicity, but
not of the order of half the system size  \cite{BJ1}. This is readily
explained by subsequent mutations after the genome duplication that
destroy the symmetry and thereby reduce the multiplicity of the
eigenvalue 1. Also, since graph duplication 
 does not change $\lambda_1$ and $\lambda_{N-1}$, the synchronization
 properties are not affected (see \cite{JJ1}).

\section{Examples of spectral plots}
\begin{figure}[h]
\begin{center}
\includegraphics[width=.5\textwidth]{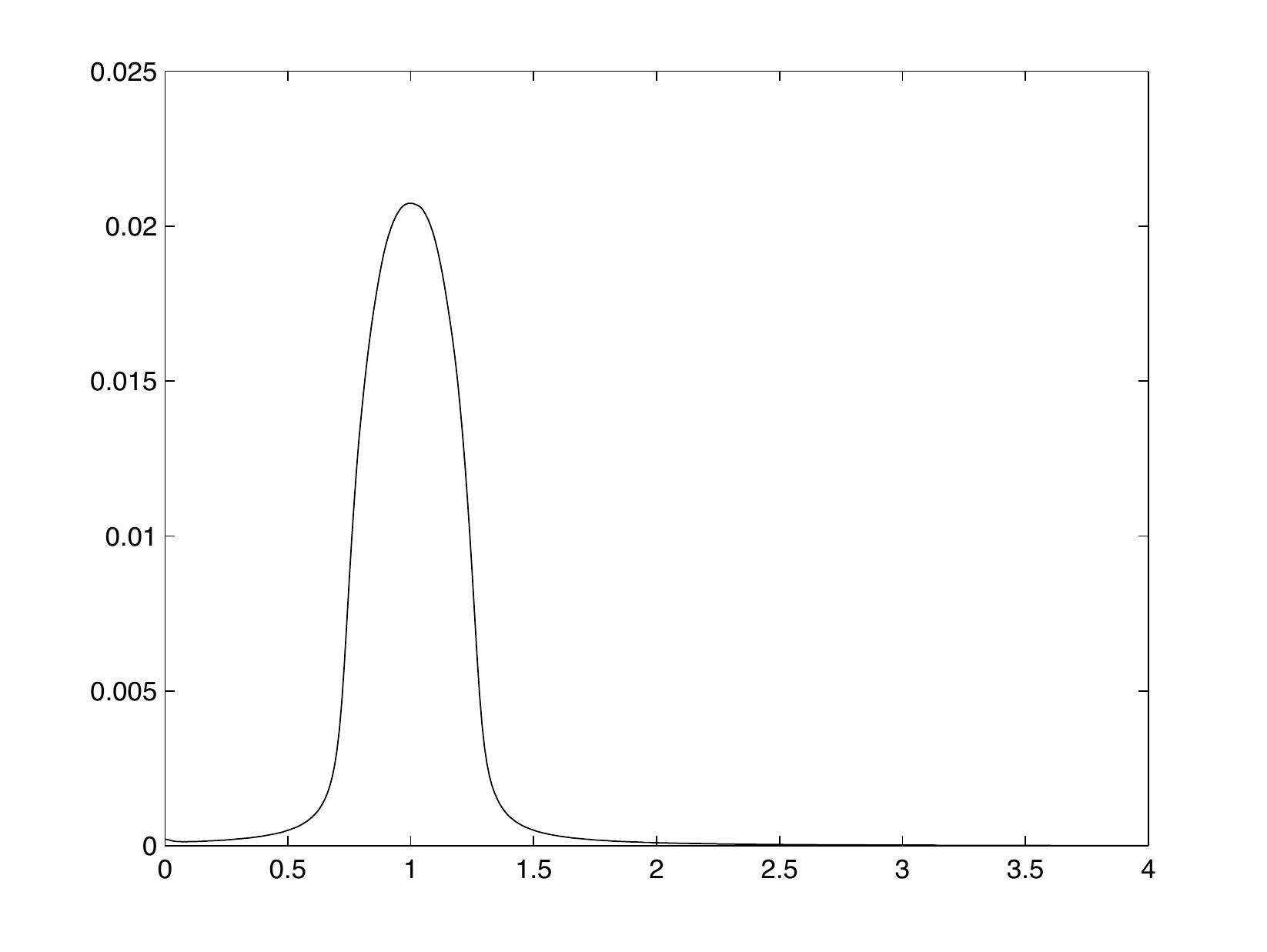}\includegraphics[width=.5\textwidth]{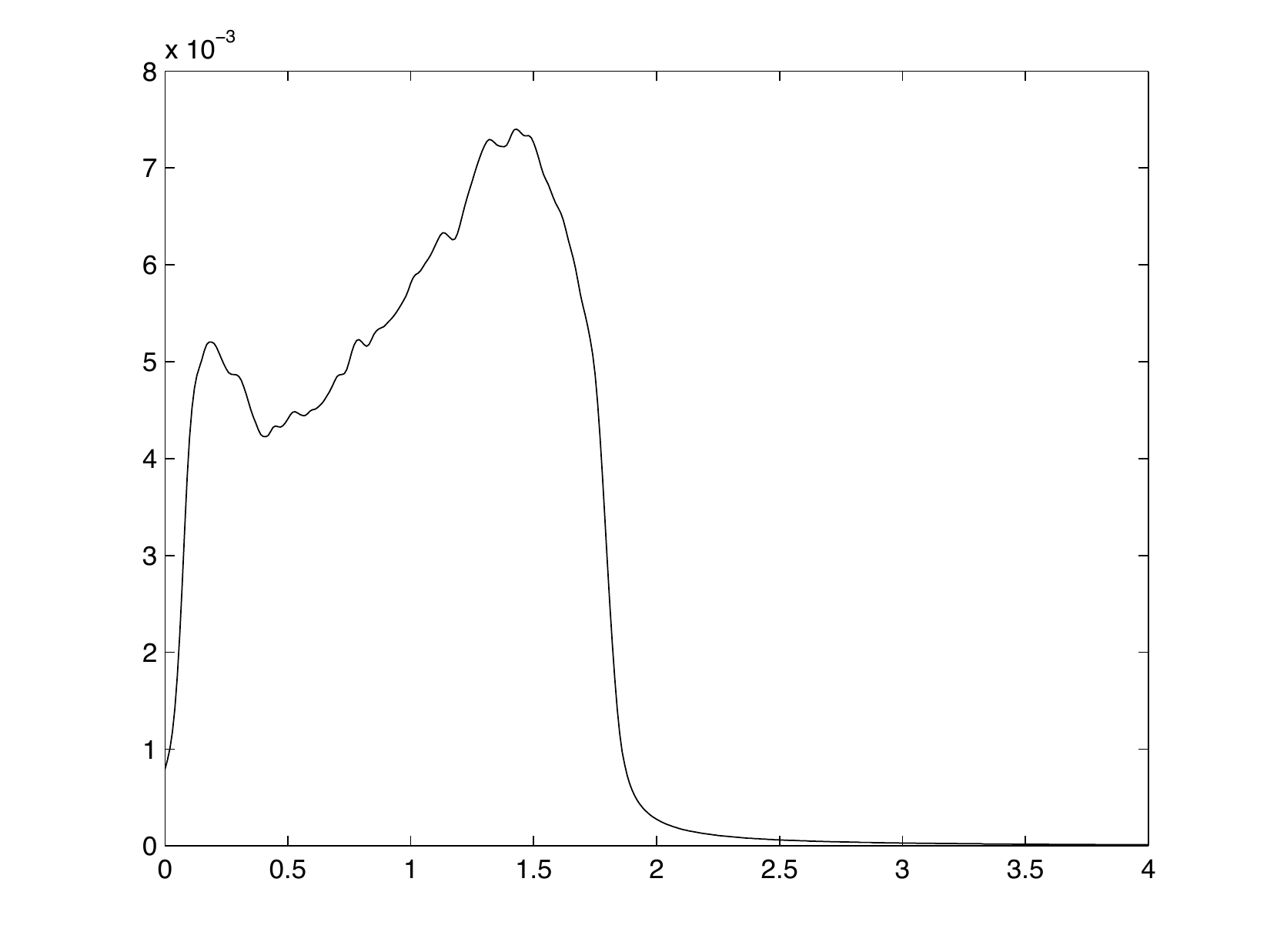}\\
(a)\hspace{.5\textwidth}(b)\\
\includegraphics[width=.5\textwidth]{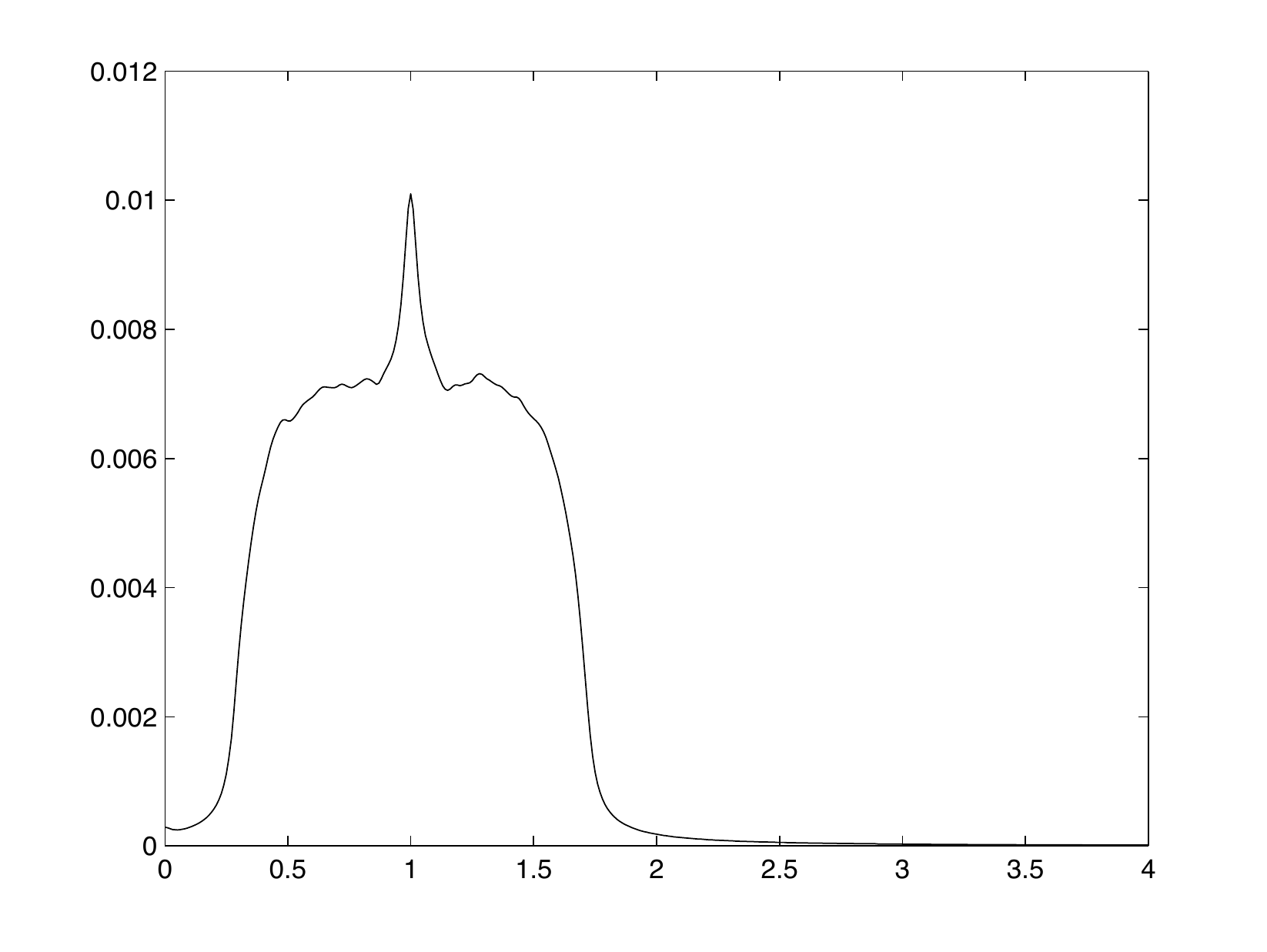}\\
(c)
\end{center}
\caption{Specral plots of generic networks. (a) Random network by Erd\"{o}s and R\'{e}nyi's model \cite{ErdosRenyi1959} with $p = 0.05$. (b) Small-world network by Watts and Stogatz model \cite{WattsStrogatz1998} (rewiring a regular ring lattice of average degree 4 with rewiring probability $0.3$). (c) Scale-free network by Albert and Barab\'{a}si model \cite{BA} ($m_0 = 5 \text{ and } m = 3 $). {\it Size of all networks is $1000$. All figures are ploted with $100$ realization}.}
\label{GeneralModelsFigs}
\end{figure}

\begin{figure}[h]
\begin{center}
\includegraphics[width=.5\textwidth]{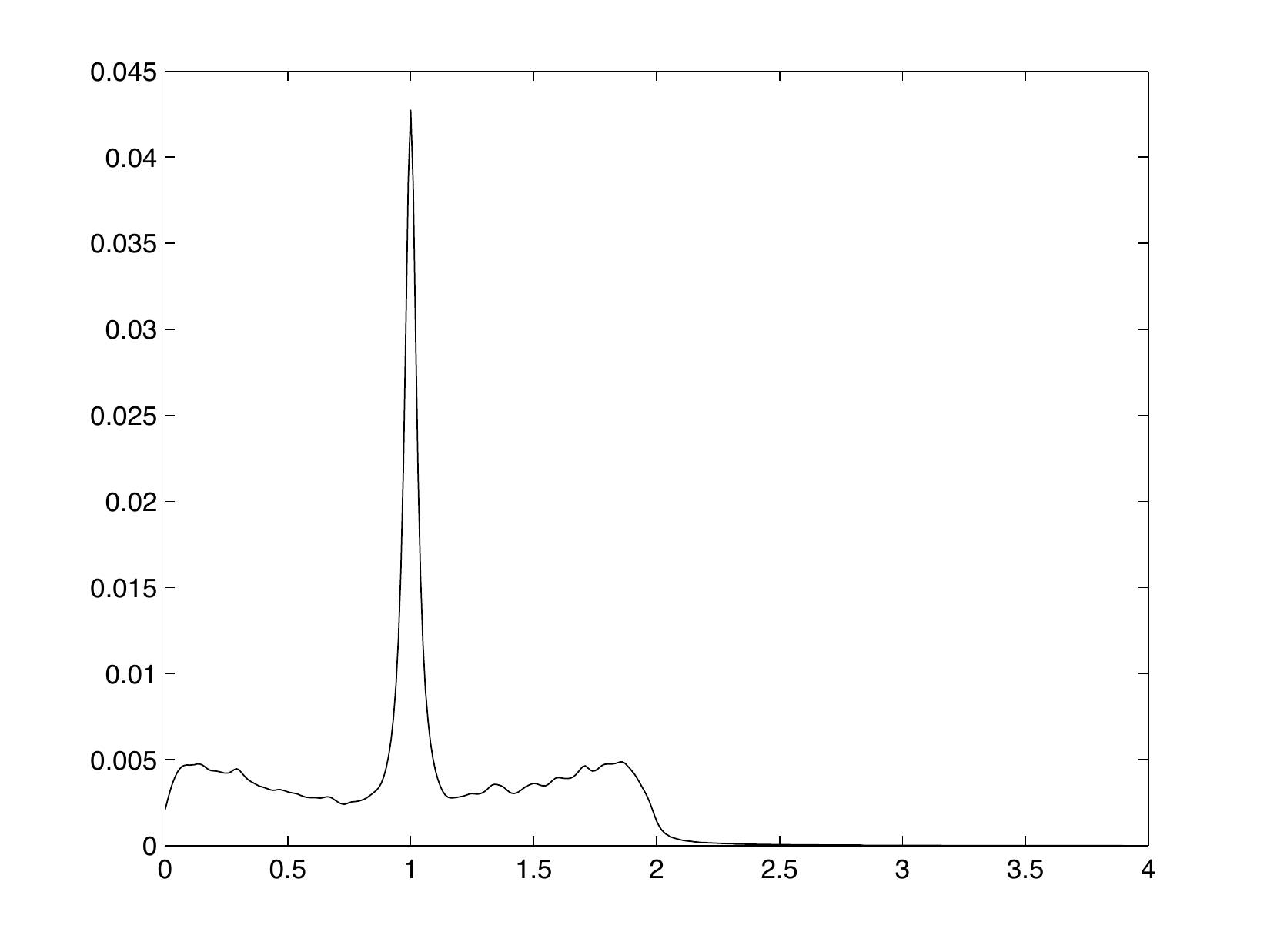}\includegraphics[width=.5\textwidth]{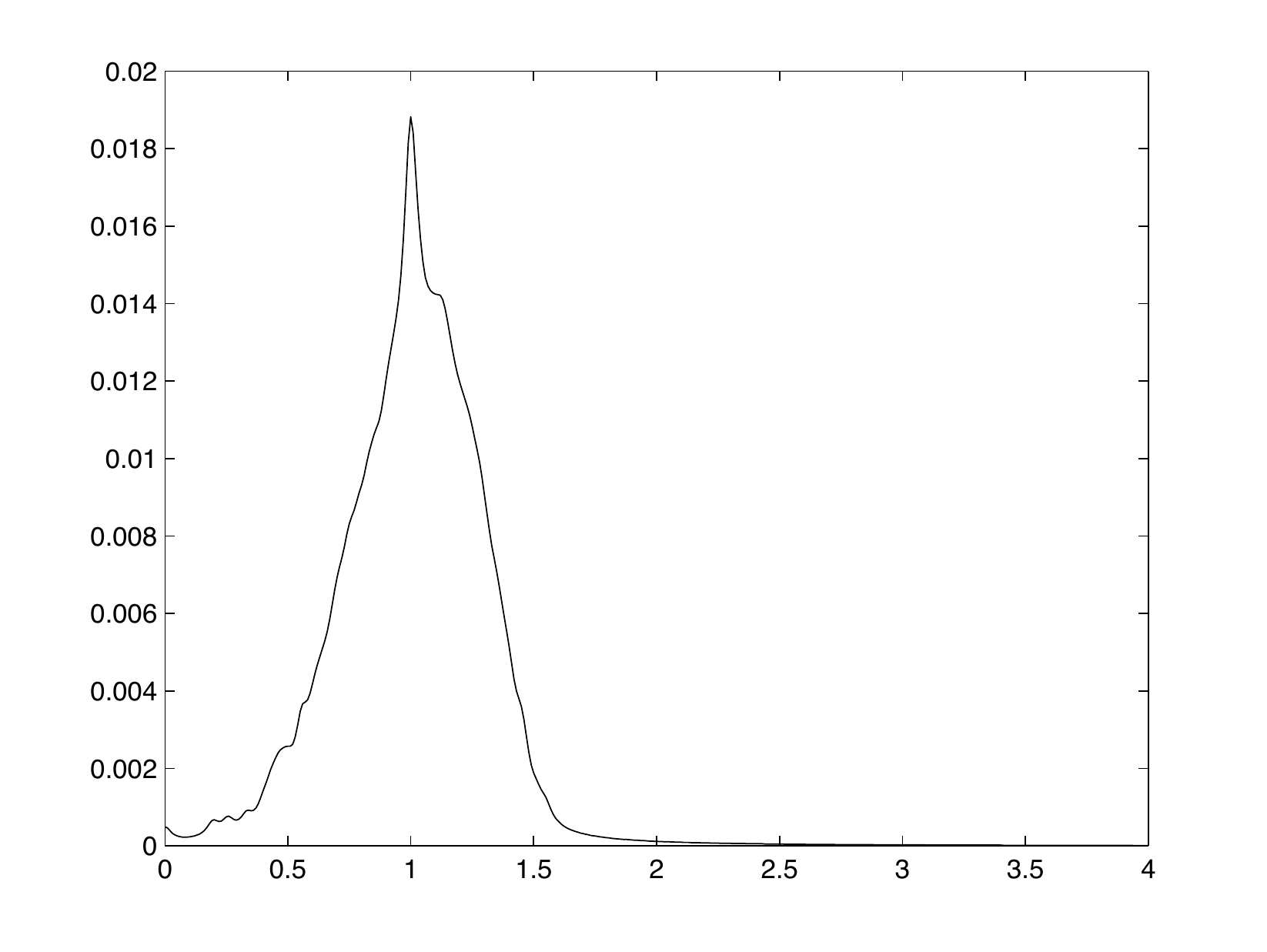}\\
(a)\hspace{.3\textwidth}(b)\\
\includegraphics[width=.5\textwidth]{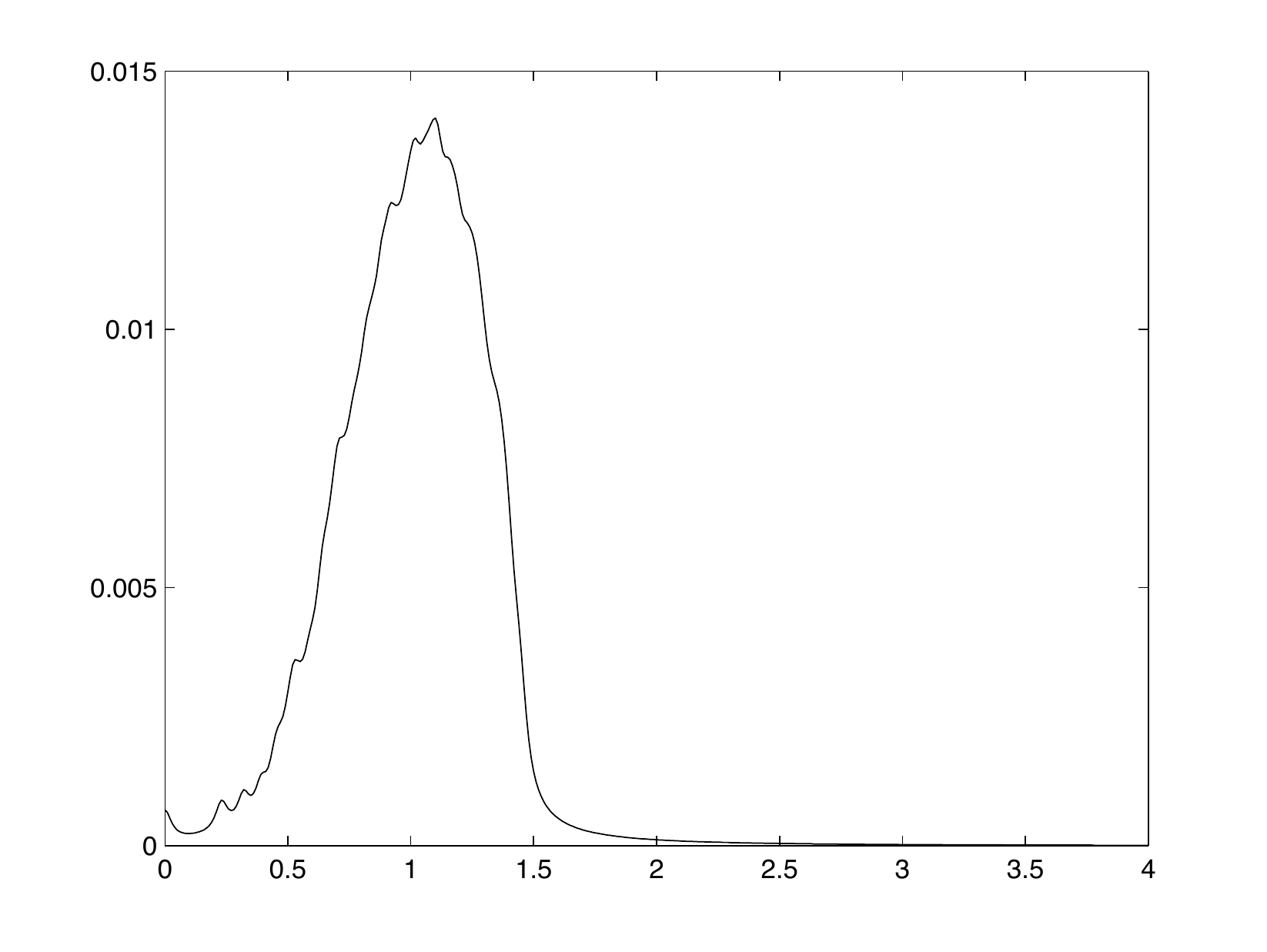}\\
(c)
\end{center}
\caption{Spectral plots of (a) protein-protein interaction network of  \textit{Saccharomyces cerevisiae} (yeast). Network size is 1458. Data downloaded from http://www.nd.edu/$\sim$networks/ and data used in \cite{JeongEtAl2001} [download date: 17th September, 2004]  (b) neuronal connectivity of {\it C. elegans}.  Size of the network is 297. Data used in \cite{WattsStrogatz1998,WhiteEtAl1986}. Data Source: http://cdg.columbia.edu/cdg/datasets/ [Download date: 18th Dec. 2006]. (c)  neuronal connectivity of {\it C. elegans} from the animal JSH, L4 male in the nerve ring and RVG regions. Network size is 190. Data source: Data is assembled by J. G. White, E. Southgate, J. N. Thomson, S. Brenner \cite{WhiteEtAl1986} and was later revisited by R. M. Durbin (Ref. http://elegans.swmed.edu/parts/ ). [Download date: 27th Sep. 2005].}
\label{PPIN_Neuro}
\end{figure}

\begin{figure}[h]
\begin{center}
\includegraphics[width=.5\textwidth]{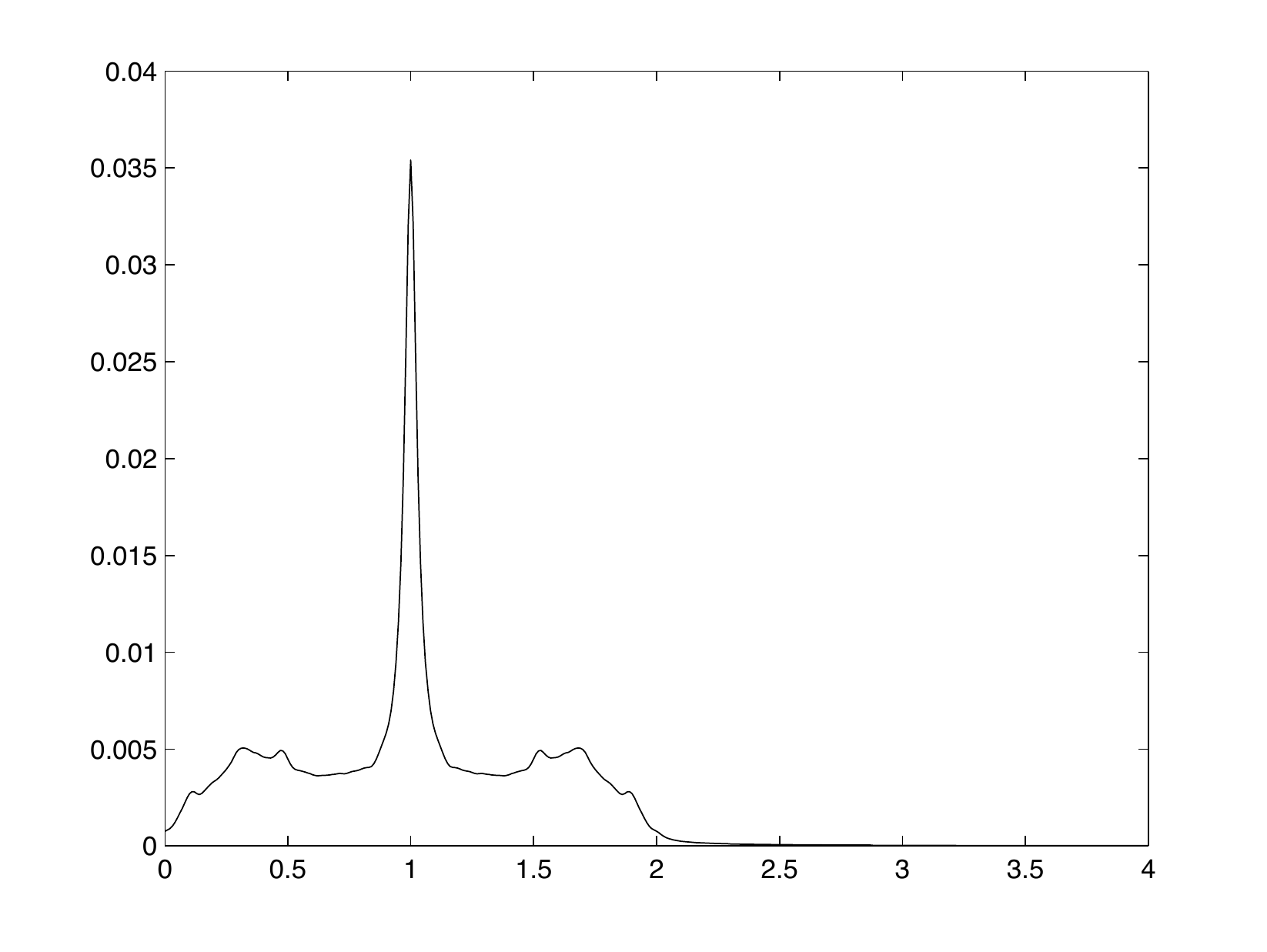}\includegraphics[width=.5\textwidth]{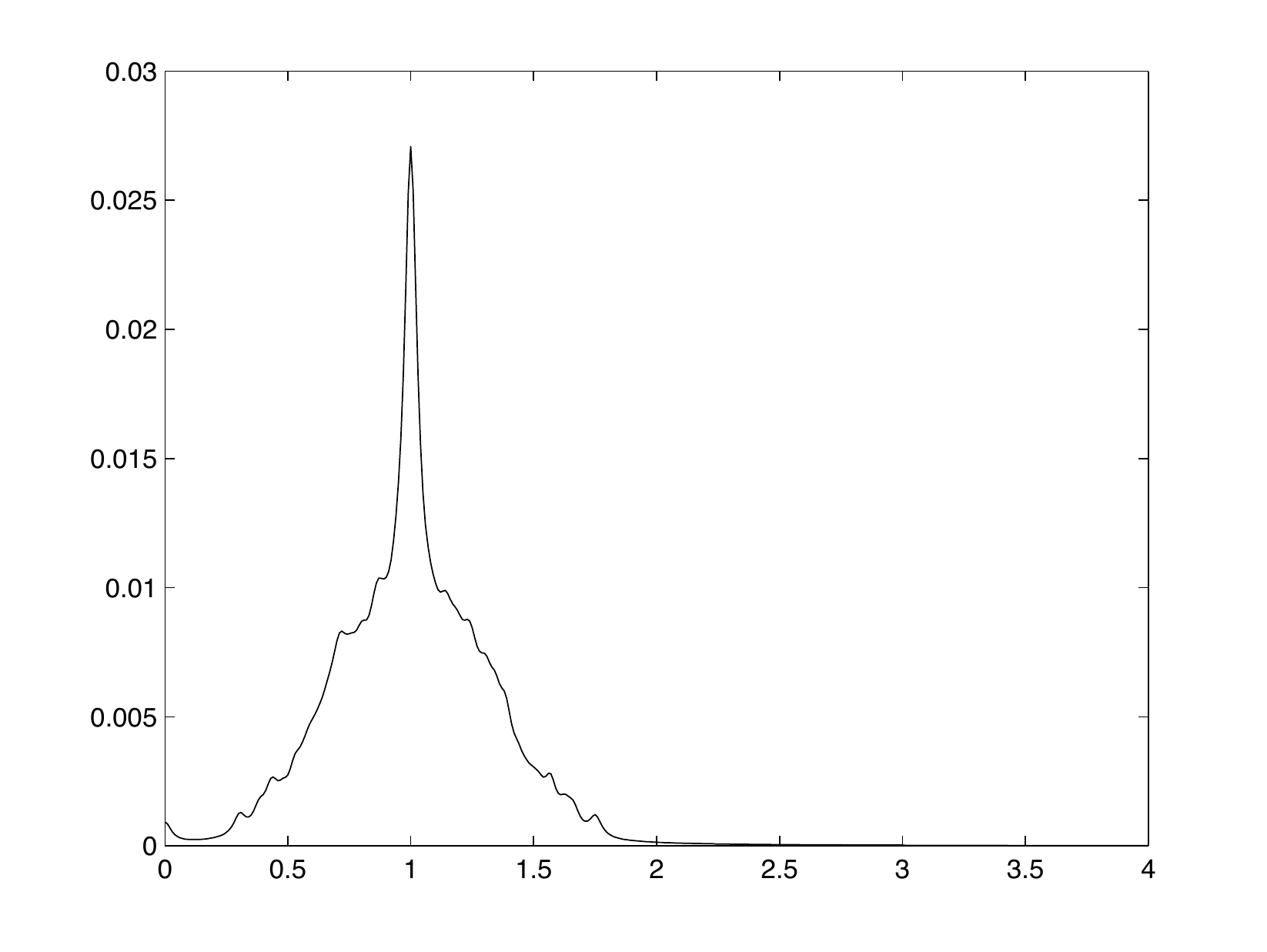}\\
(a)\hspace{.3\textwidth}(b)\\
\includegraphics[width=.5\textwidth]{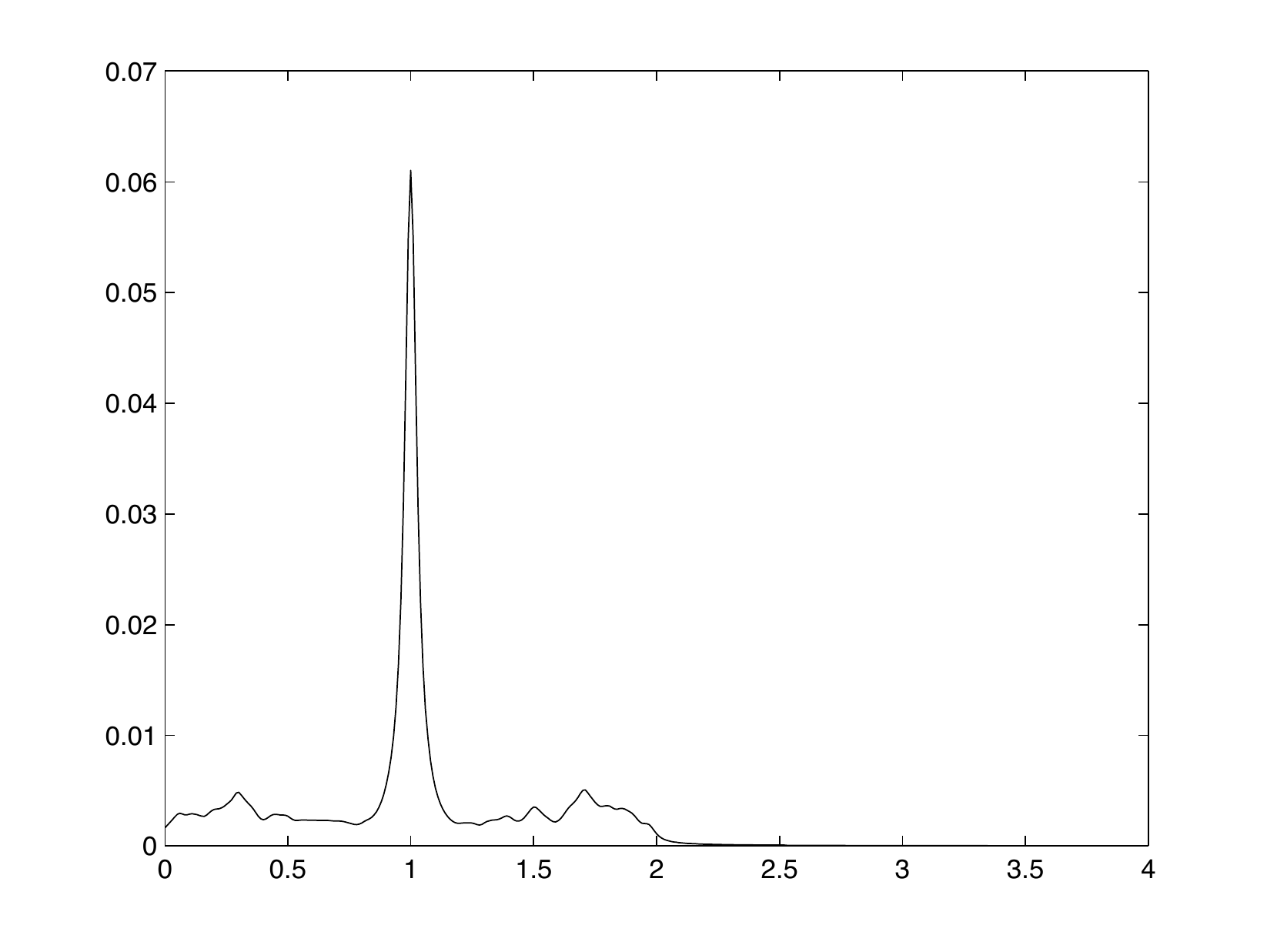}\\
(c)
\end{center}
\caption{Spectral plots of (a) metabolic network of {\it A. pernix}. Network size is 490. Here nodes are substrates, enzymes and intermediate complexes. Data used in \cite{JeongEtAl2000}. Data Source: http://www.nd.edu/$\sim$networks/resources.htm/. [Download date: 22nd Nov. 2004].  (b) food-web from "Ythan estuary". Size of the network is 135. Data downloaded from http://www.cosin.org/. [Download Date 21st December, 2006]. (c)  transcription network of {\it E. coli}. Size of the network is 328. Data source: Data published by Uri Alon (http://www.weizmann.ac.il/mcb/UriAlon/ ). [Download date: 13th Oct. 2004]. Data used in \cite{MiloEtAl2002,Shen-OrrEtAl2002}.}
\label{Meta_FWeb_Trans}
\end{figure}

We now exhibit spectral plots of different formal and biological networks. We convolve the eigenvalues with a Lorentz kernel, that is, we plot the graph of the function
$$f(x)=\sum_{\lambda_j}\frac{\gamma}{(\lambda_j-x)^2+\gamma^2}$$
where the $\lambda_j$ are the eigenvalues and we choose the parameter
value $\gamma=.03$.\footnote{In fact, we could as well take some other
  kernel here; the general formula is $f(x)=\sum_{\lambda_j}\int
  k(x,\lambda)\delta(\lambda-\lambda_j)d\lambda$ where $k(x,\lambda)$
  is some kernel function. As an alternative to the Lorentz kernel, we
  could also take, e.g., a Gaussian kernel, or a piecewise constant
  kernel $k(x,\lambda)=\frac{1}{2\gamma}$ if $|x-\lambda|\le \gamma$
  and 0 else. The shape of the kernel is less important here than a
  careful choice of the parameter $\gamma$. For small $\gamma$, the
  plot obscures the global features, while for large $\gamma$, the
  details become too blurred.}
\\
In Figure \ref{GeneralModelsFigs}, we see an Erd\"os-R\'enyi random network, a Strogatz-Watts small-world network, and a Barab\'asi-Albert scale-free network. Each of these types has its very distinct shape, and this is not affected by varying the parameters underlying the construction schemes. In Figure \ref{PPIN_Neuro}, we then see a protein-protein interaction network and two neurobiological networks, and in Figure \ref{Meta_FWeb_Trans}, we have a metabolic network, a food-web, and a transcription network. These are just examples, and choosing other examples from the same category yields very similar shapes. As we directly see, however, shapes of spectral plots for networks from different biological realms are very different from each other and from the formal networks, even though those have been suggested to capture important aspects of  biological networks. Clearly, this indicates that for analysing biological networks, it does not suffice to rely on some generic formal construction scheme. Rather, one needs to analyse the specific aspects of specific biological realms through formal methods that are sufficiently rich to capture the essential qualitative features of that biological domain. In this paper, we have proposed spectral analysis as such a method.

\begin{thebibliography}{999}

\bibitem{AB}
R. Albert, A.-L. Barab{\'a}si, Statistical mechanics of complex networks,
  Reviews of Modern Physics 74, 2002, 47--97.



\bibitem{ABJ1}
F.M. Atay, T.B{\i}y{\i}ko{\u g}lu, J.Jost, Synchronization of networks with
  prescribed degree distributions, {IEEE} Trans.~Circuits and Systems
  I 53~(1), 2006, 92--98.

\bibitem{ABJ2}
F.M. Atay, T.B{\i}y{\i}ko{\u g}lu, J.Jost, Network synchronization:
Spectral versus statistical properties, Phys.D, to appear

\bibitem{AJW}
F.M. Atay, J. Jost, A. Wende, Delays, connection topology, and synchronization
  of coupled chaotic maps, Phys. Rev. Lett. 92~(14), 2004, 144101.

\bi{BJ1} A.Banerjee, J.Jost, Laplacian spectrum and protein-protein
interaction networks, e-print: arXiv:0705.3373v1


\bi{BJ2} A.Banerjee, J.Jost, On the spectrum of the
normalized graph Laplacian, e-print: arXiv:0705.3772v1


\bibitem{BA}
A.-L. Barab{\'a}si, R.~A. Albert, Emergence of scaling in random networks,
  Science 286, 1999, 509--512.



\bi{BiLeSt} T.B{\i}y{\i}ko{\u{g}}lu, J.Leydold, P.Stadler, Laplacian eigenvectors
of graphs, Springer LNM, to appear

\bi{Bol} B.Bolob\'as, Modern graph theory, Springer, 1998

\bi{Chung} F.Chung, Spectral graph theory, AMS, 1997

\bibitem{DoMe}
S.N. Dorogovtsev, J.F.F. Mendes, Evolution of Networks, Oxford, 2003.

\bibitem{DLS}
G.Gladwell, E.Davies, J.Leydold, and P.Stadler,
 Discrete nodal domain theorems,
 Lin.Alg.Appl.336, 2001, 51-60


\bi{ErdosRenyi1959} P. Erd\"{o}s, A. R\'{e}nyi,
On random graphs,
Publicationes Mathematicae Debrecen, 6, 1959, 290-297


\bi{GoRo} C.Godsil, G.Royle, Algebraic graph theory, Springer, 2001

\bi{IpMi} M. Ipsen, A. S. Mikhailov,
  Evolutionary reconstruction of networks,
  Phys. Rev. E 66(4),
  2002


\bi{JeongEtAl2000} H. Jeong and B. Tombor and R. Albert and Z. N. Oltval and A. L. Barab\'{a}si,
The Large-Scale Organization of Metabolic Networks,
Nature, 407(6804), 2000, 651-654

\bi{JeongEtAl2001} H. Jeong, S. P. Mason, A. L. Barab\'{a}si, Z. N. Oltvai,
   Lethality and Centrality in Protein Networks,
   Nature, 411(6833), 2001, 41-42


\bi{J1} J. Jost, Mathematical methods in biology and neurobiology,
monograph, to appear

\bi{J2} J. Jost, in: J.F.Feng, J.Jost, M.P.Qian (eds.), Networks: from
biology to theory, Springer, 2007

\bi{JJ1} J. Jost, M. P. Joy,
  Spectral properties and synchronization in coupled map lattices,
  Phys.Rev.E 65(1), 2002, 016201
 
\bi{Merris}
  R. Merris, Laplacian matrices of graphs -- a survey,
   Lin. Alg.  Appl.198, 1994, 143-176


\bi{MiloEtAl2002} R. Milo and S. Shen-Orr and S. Itzkovitz and N. Kashtan and D. Chklovskii and U. Alon,
Network Motifs: Simple Building Blocks of Complex Networks,
Science, 298(5594), 2002, 824-827


\bi{Mohar} B. Mohar,
 Some applications of Laplace eigenvalues of graphs, in: G.Hahn,
 G.Sabidussi (eds.),
 Graph symmetry: Algebraic methods and applications, pp. 227-277,
 Springer, 1997

\bibitem{New}
M. Newman, The structure and function of complex networks, {SIAM} Review
  45, 2003, 167--256
\bi{Ohno1970} S.Ohno, Evolution by Gene Duplication, Springer, 1970


\bibitem{PeCa}
L.M. Pecora, T.L. Carroll, Synchronization in chaotic systems, Phys. Rev.
  Lett. 64, 1990, 821--824
\bibitem{PRK}
A.Pikovsky, M.Rosenblum, J.Kurths, Synchronization -- {A} Universal Concept
  in Nonlinear Science, Cambridge University Press, Cambridge, 2001
 

\bi{Shen-OrrEtAl2002} S. S. Shen-Orr and R. Milo and S. Mangan and U. Alon,
Network Motifs in the Transcriptional Regulation Network of Escherichia Coli,
Nature Genetics, 31(1), 2002, 64-68


\bi{Wagner1994} A. Wagner,
Evolution of Gene Networks by Gene Duplications - a Mathematical-Model and Its Implications On Genome Organization,
Proc. Nat. Acad. Sciences USA  91(10), 1994, 4387-4391


\bi{WattsStrogatz1998} D. J. Watts, S. H. Strogatz,
Collective Dynamics of 'Small-World' Networks,
Nature, 393(6684), 1998, 440-442

\bi{WhiteEtAl1986} J. G. White and E. Southgate and J. N. Thomson and S. Brenner,
The Structure of the Nervous-System of the Nematode Caenorhabditis-Elegans,
Philosophical Transactions of the Royal Society of London Series B-Biological Sciences, 314(1165), 1986, 1-340


\bi{WolfeShields1997} K. H. Wolfe, D. C. Shields,
Molecular Evidence for an Ancient Duplication of the Entire Yeast Genome,
Nature, 387(6634), 1997, 708-713

\bi{ZW} P.Zhu, R.Wilson, A study of graph spectra for comparing
graphs, http://cms.brookes.ac.uk/staff/PhilipTorr/BMVC2005/papers/paper-57-162.html

\end{thebibliography}
\end{document}